\documentclass[
reprint,
onecolumn,
pra,
notitlepage
]{revtex4-1}

\usepackage{dcolumn}
\usepackage{bm}
\usepackage{graphicx} 
\usepackage{float}

\usepackage{multirow}
\usepackage{booktabs}
\usepackage{array} 
\usepackage{paralist} 
\usepackage{verbatim} 
\usepackage{subfigure} 
\usepackage{rotating}
\usepackage{fancyhdr} 
\usepackage{threeparttablex}
\usepackage{amsmath}

\graphicspath{{Figures/}}

\usepackage{tikz-inet}	
	\usetikzlibrary{shapes,snakes,arrows,circuits.logic.US}
	\usepackage{circuitikz}
	\usetikzlibrary{backgrounds,fit,decorations.pathreplacing} 
	\tikzstyle{branch}=[fill,shape=circle,minimum size=3pt,inner sep=0pt]
	
	\usepackage{xcolor}
\newcommand{\ket}[1]{\ensuremath{\left|#1\right\rangle}} 

\begin{document}
\title{Quantum Eigenvalue Estimation for Irreducible Non-negative Matrices} 
\author{Anmer~Daskin}\email{email:anmerdaskin@yahoo.com}
\affiliation{Department of Computer Engineering, Istanbul Medeniyet University, Uskudar, Istanbul, Turkey}

\begin{abstract}

Quantum phase estimation algorithm has been  successfully adapted as a sub frame of many other algorithms applied to a wide variety of applications in different fields. However, the requirement of a good approximate eigenvector given as an input to the algorithm hinders the application of the algorithm to the problems where we do not have any prior knowledge about the eigenvector.

In this paper, we show that  the principal eigenvalue of an irreducible non-negative operator can be determined by using an equal superposition initial state in the phase estimation algorithm. This removes the necessity of the existence of  an initial good approximate eigenvector. Moreover,  we show that the success probability of the algorithm is related to the closeness of the operator to a stochastic matrix. Therefore, we draw an estimate for the success probability by using the variance of the column sums of the operator.
This provides a priori information which can be used to know the success probability of the algorithm beforehand for the non-negative matrices and apply the algorithm only in cases when the estimated probability reasonably high. Finally, we discuss the possible applications and show the results for random symmetric matrices and 3-local Hamiltonians  with non-negative off-diagonal elements.
\end{abstract}

\maketitle

\markboth{A Modified Phase Estimation Algorithm}{A Modified Phase Estimation Algorithm}

\section{Introduction}

In the recent decades, there have been many quantum algorithms proposed for  the problems inefficient to solve on classical computers. The quantum phase estimation algorithm \cite{Abrams,Kitaev} is a leading illustration of such algorithms. It is used for quantum simulations to estimate the eigenenergy corresponding to a given approximate eigenvector of
the unitary evolution operator of a quantum Hamiltonian. Furthermore,
with different settings, it has been  adapted as a sub frame of many quantum algorithms applied to wide variety of applications in different fields (see the review article ref.\cite{Georgescu2014quantum} and the references therein).
However, the success of the algorithm partly depends on the pre-existing approximate eigenvector. Thus, it may fail to output the right eigenvalue in the cases where the approximation to the corresponding eigenvector is not good enough.  This hinders  the uses of the algorithm when there  is not enough prior knowledge to procure a good approximation to the eigenvector. 

A symmetric matrix is called non-negative if all of its elements are greater than or equal to zero. It is irreducible if the corresponding graph is strongly connected or if it is not reducible to a block-diagonal form by any column-row permutation. Due to Perron-Frobenius theorem \cite{Meyer2000}, an irreducible non-negative matrix have a positive eigenvector (all elements are positive) with an associated positive  principal eigenvalue whose magnitude is greater than the rest of the eigenvalues. These matrices have been studied extensively \cite{Bapat1997nonnegative,Berman1979nonnegative}, the distribution of the coefficients of their principal eigenvectors have been related to the matrix elements \cite{Minc1970maximal}, and the sum of the coefficients have been shown to be related to the number of walks in molecular graphs \cite{Gutman2001}.

The phase estimation algorithm (PEA) mainly uses two quantum registers:$\ket{reg1}$ initially set to zero state and $\ket{reg_2}$ holding an initial approximate eigenvector.
In this article, we show that for irreducible non-negative matrices, one can obtain the principal eigenvalue by using an equal superposition input state  in PEA:
i.e., instead of an approximate eigenvector, initial value of \ket{reg_2} is set to \ket{\bf 0} and then put into equal superposition state by applying Hadamard gates. In the output of the algorithm, this generates each one of the eigenvalues with the probability determined by the normalized sum of the coefficients of the associated eigenvector. 
In addition, because  all eigenvectors but the principal one include positive-negative elements, we show that in most of the random cases the probability to see principal eigenvalue in the output is much larger than the others.    
In also some cases, we show that first applying Hadamard gates to the second register in the output (or measuring second register in the Hadamard basis) and then measuring the first register when the measurement outcome of the second register is  in \ket{0\dots 0} state    increase the success probability further in the output.

Eigenvector-coefficients of stochastic matrices sum to zero for all but the principal eigenvector. Therefore, for these matrices one can generate the eigenvalue in the phase estimation algorithm with probability one \cite{Daskin2014}. Since any given symmetric irreducible non-negative matrix can be converted into a stochastic matrix by a diagonal scaling matrix; using the closeness of the matrix to a stochastic matrix, we also show that the success probability of the algorithm can be predicted beforehand. 
We finally compare the estimated success probabilities with the computed ones for random symmetric matrices  and 3-local Hamiltonians of different dimensions with non-negative off-diagonal elements.

In the following sections, we shall first discuss PEA in detailed steps, then draw the estimates for the success probability and finally show the possible applications of the algorithm.

\section{Quantum Phase Estimation Algorithm}
The phase estimation algorithm (PEA) in general sense finds the value of $\phi_j$ for a given approximate eigenvector $\ket{\mu_j}$ in the eigenvalue equation  $U\ket{\mu_j}=e^{i\phi_j}\ket{\mu_j}$. 
The algorithm mainly uses two quantum registers: Viz. $\ket{reg1}$ initially set to zero state and $\ket{reg_2}$ holding an approximate eigenstate of
  a unitary matrix $U$. The first operation in the algorithm puts \ket{reg1} into the equal superposition state. In this setting, 
 a sequence of operators, $U^{2^j}$, controlled by the $j$th qubit
of $\ket{reg1}$ are applied to $\ket{reg_2}$. Here, $j=1 \dots m$ and $m$ is the number of qubits in \ket{reg1} and also determines the precision of the output. These sequential operations generate the quantum Fourier transform $(QFT)$ of the
phase on $\ket{reg1}$. Therefore, the application of the inverse quantum Fourier transform $(QFT^\dagger)$ turns  the value of $\ket{reg1}$  into the binary value of the phase. Consequently, one measures \ket{reg1} to obtain the phase.  Here, if the unitary operator $U$ is the time evolution operator of a quantum Hamiltonian $\mathcal{H}$, i.e. $U=e^{i\mathcal{H}}$; then one also obtains the eigenenergy of that Hamiltonian.

For a symmetric irreducible non-negative matrix $\mathcal{H}$ of order $2^n$ with ordered  eigenvalues as $\phi_1\geq \dots \geq \phi_{2^n}$ and associated eigenvectors $\ket{\mu_1} \dots \ket{\mu_{2^n}}$, it is known that $\ket{\mu_1}$ is the only eigenvector with all positive coefficients.   Assume unitary operator $U=e^{i\mathcal{H}}$  and its powers $U^{2^j}$ with $j=0 \dots m$ are readily available for PEA. To estimate the value of $\phi_1$ on the first register,  in our setting, we simply modify the conventional  phase estimation algorithm in the following way: 
\begin{itemize}
\item Instead of an approximate eigenvector, initial value of \ket{reg_2} is set to \ket{\bf 0} and then put into equal superposition state by applying Hadamard gates.  
\item (\textbf{optional}) In the output, we change the basis of the second register  by applying the Hadamard gates. Then, if the measurement of the second register is equal to \ket{\bf 0} state, then the principal eigenvalue is estimated  on the first register. This modification increases the success probability of the algorithm further when the probability of measuring the principal eigenvalue is higher then the other eigenvalues.
\end{itemize} 
 
In the following subsection,  we shall describe the phase estimation algorithm in steps, also shown in Fig.\ref{fig:peageneral}, to show how the above modifications have an impact on the algorithm.
\subsection{Steps of the Algorithm} 
Here, the algorithm is described in details by showing how the quantum state changes in each step:
\begin{itemize}
\item The system is prepared in a way so that it includes two register: viz., \ket{reg_1} and \ket{reg_2} with $m$ and $n$ number of qubits, respectively.
\item Both registers are initialized into zero state: 
\ket{\psi_0}=\ket{reg_1}\ket{reg_2}=\ket{\bf 0}\ket{\bf 0}.
\item Then, the Hadamard operators are applied to both registers:
\begin{equation}
\ket{\psi_1}=(H^{\otimes m} \otimes H^{\otimes n}) \ket{\bf 0} \ket{\bf 0}
=\frac{1}{\sqrt{2^{n+m}}}\sum_{x=0}^{2^{m+n}-1}\ket{\bf x}
\end{equation}

\item As in the customary phase estimation algorithm;
 the unitary evolution operators $U^{2^j}$ controlled by the $j$th qubit of \ket{reg_1} are applied to \ket{reg_2},
 and finally the inverse of the quantum Fourier transform ($QFT^{\dagger}$) is applied to \ket{reg1}.

At this point of the algorithm, we have a quantum state in which \ket{reg_1} and \ket{reg_2} hold the superposition of the eigenvalues and the associated eigenvectors, respectively: 
\begin{equation}
\ket{\psi_2}=\sum_{j=0}^{N-1} \alpha_j \ket{\phi_j}\ket{\mu_j}.
\end{equation}
 Here, the amplitude $\alpha_j$ is related to the angle between the equal superposition state and the eigenvector \ket{\mu_j} and can be described as the normalized sum of the coefficients of the eigenvectors since \ket{reg_2} was $\frac{1}{\sqrt{2^n}}\sum_{x=0}^{2^{n}-1}\ket{\bf x}$ at the beginning:
 \begin{equation}
 \label{eq:normalizedsum}
 \alpha_j=1/\sqrt{2^n}\sum_{i=1}^{2^n}{\mu_{ji}}
 \end{equation}
 If \ket{reg_2} is written in the Hadamard basis,
 $\ket{+\dots ++} +
\ket{+\dots +-} +\dots 
+\ket{-\dots --}$, where $\ket{\pm}=1/\sqrt(2)(\ket{0}\pm\ket{1})$;
 then  the following quantum state is obtained:
\begin{equation}
\begin{split}
\label{EqH}
\ket{\psi_{2}}& = \sum_{j=1}^{2^{n}}\beta_{1j}\ket{ \phi_j}\ket{+\dots +}
\\ &  
+ \dots +\sum_{j=1}^{2^{n}}\beta_{2^nj}\ket{ \phi_j}\ket{-\dots -}  
\end{split}
\end{equation}
where $\beta_{ij}$s are new coefficients.
Now, since there is only one eigenvector, viz. \ket{\mu_1},  with  all positive real elements,  in the Hadamard basis, we can expect this positive eigenvector to be the closest state to \ket{++\dots ++}.

\item 
To revert the above state back to the standard basis, the Hadamard operator $H^{\otimes n}$ is again applied to \ket{reg_2}:
\begin{equation}
\begin{split}
\label{EqFinal}
\ket{\psi_3}& =(I\otimes H^{\otimes n})\ket{\psi_2}\\
 & = \sum_{j=1}^{2^{n}}\beta_{1j}\ket{ \phi_j}\ket{0\dots 0}
+ \dots +\sum_{j=1}^{2^n}\beta_{2^nj}\ket{ \phi_j}\ket{1\dots 1}  
\end{split}
\end{equation}
 
\item  \ket{reg_2} is measured  in the standard basis. As a result, for \ket{reg_2}= \ket{0\dots 00}, the system collapses to the state where  the phase associated with the positive eigenvector is  expected to be highly dominant in the first register.

\item In the final step, the first register is measured to obtain the phase $\phi_1$ and get the eigenvalue of $\mathcal{H}$. These steps also drawn in Fig.\ref{fig:peageneral}.

\begin{figure}[h]
\centering
\includegraphics[width=0.5\textwidth]{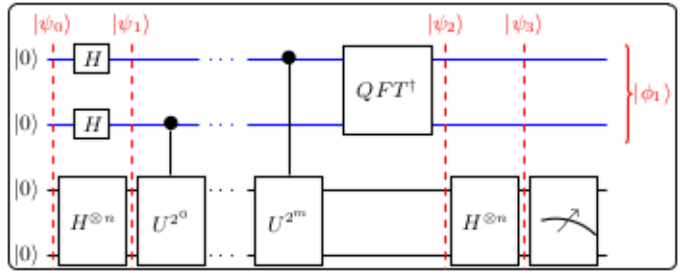}
\caption{Circuit design for the modified phase estimation algorithm. In the end, note that the application of the Hadamard gates to the second register and the measurement on this register is optional.
  }
\label{fig:peageneral}
\end{figure}
\end{itemize}
\subsection{The Success Probability}
It is easy to see that the success probability of the algorithm without the optional Hadamard gates on \ket{reg_2} in the output is determined from the normalized sum of the coefficients of the eigenvectors given in Eq.(\ref{eq:normalizedsum}). Therefore, the success probability for the dominant eigenvalue is $\alpha_1^2$ (note that the coefficients of \ket{\mu_1} are all positive, hence,  $\alpha_1^2=|\alpha_1|^2$.).

With the Hadamard gates in the output, we first find the probability to measure \ket{reg_2} in \ket{\bf 0} state and then find the probability to see the dominant eigenvalue in \ket{reg_1} knowing $\ket{reg_2}=\ket{\bf 0}$:
  As obvious in Eq.(\ref{EqFinal}), the success probability of getting \ket{\bf 0} is $\left(\sum_{j=1}^{2^n}{\beta_{1j}}\right)^2$. In addition, after measuring \ket{\bf 0} on \ket{reg_2}, the probability to measure  $\ket{\phi_1}$ on \ket{reg_1} is 
$|\beta_{11}|^2$. 
Since we apply an equal superposition as an input, the component of an eigenvector in this direction determines the probability to get the corresponding eigenvalue on \ket{reg_1}.  More formally, after the application of $QFT^\dagger$, we get $\ket{\psi_2}=\sum_{j=0}^{N-1} \alpha_j \ket{\phi_j}\ket{\mu_j}$, where $\alpha_j=1/\sqrt{2^n}\sum_{i=1}^{2^n}{\mu_{ji}}$. 
If  \ket{reg_2} of the state in Eq.(\ref{EqH}) is measured in the Hadamard basis, the probability to see \ket{+\dots+} is the sum of the amount of the components of the eigenvectors in the direction of the initial vector: 
\begin{equation}
P_{reg_2}=\sum_{j=1}^{2^n}|\alpha_j|^4\geq \frac{1}{2^n}
\end{equation}
This is also equal to $\sum_{j=1}^{2^n}{\beta_{1j}}$, i.e. the probability of measuring \ket{reg_2} in \ket{0\dots0} state at the end. $P_{reg_2}$ takes the smallest value only when all $|\alpha_j|$s are equal to  $1/\sqrt{2^n}$.  Moreover, when \ket{reg_2}=\ket{0\dots 0}, the probability to see $\ket{\phi_1}$ on \ket{reg_1} is: 
\begin{equation}
P_{reg_1}=|\beta_{11}|^2=\frac{\alpha_1^4}{P_{reg_2}}. 
\end{equation}

Although we have drawn the equalities for probabilities, without knowing the eigenvectors, it is not possible to compute exact $|\alpha_j|$s and so $P_{reg_1}$ and $P_{reg_2}$. 
Therefore, we shall try to estimate them: Since $\alpha_1$ is the normalized sum of the vector elements, it is easy to see that $\alpha_1\geq 1/\sqrt{2^n}$, where the equality occurs only when the magnitude of the eigenvectors are the columns of the identity matrix. A similar observation is also made in ref.\cite{Daskin2014} where the principal eigenvector is found for a given eigenvalue equal to 1. 

Here, we will relate the matrix $\mathcal{H}$ to a stochastic matrix in order to develop some intuition for the estimation of the probabilities. It is known that stochastic matrices have only one eigenvector with coefficients summing to a nonzero value.
A matrix can be made stochastic by a column wise scaling $\mathcal{HD}$, where $\mathcal{D}$ is a diagonal matrix whose diagonal elements are the inverses of the sums of the columns of $\mathcal{H}$.  The closeness of the matrix $\mathcal{H}$ to $\mathcal{HD}$ may provide a prediction to estimate the success behavior of the phase estimation algorithm with an initial superposition state.
The closeness of $\mathcal{H}$ to $\mathcal{HD}$ can be defined by  $||\mathcal{H}-\mathcal{HD}||$ where $||\cdot||$ is any matrix norm. This also defines a perturbation error. If we normalize this error term by $||\mathcal{H}||$, we get the relative error:
\begin{equation}
\epsilon_1=\frac{||\mathcal{H}-\mathcal{HD}||}{||\mathcal{H}||}\leq ||\mathcal{I}-\mathcal{D}||,
\end{equation}
where $\mathcal{I}$ is an identity matrix. We can also look at the relative error in the inverses:
\begin{equation}
\epsilon_1=\frac{||\mathcal{H}^{-1}-\mathcal{D}^{-1}\mathcal{H}^{-1}||}{||\mathcal{H}^{-1}||}\leq ||\mathcal{I}-\mathcal{D}^{-1}||.
\end{equation}

Since the matrix $\mathcal{D}$ also changes the eigenvector elements;  when the variance of the diagonal elements of $\mathcal{D}$ is small, we can expect the eigenvectors of $\mathcal{H}$ to show the similar behaviors as of $\mathcal{HD}$ and have one eigenvector whose elements sum to a much greater value than the others. When $\mathcal{H}=\mathcal{HD}$ or $\mathcal{D=I}$, left and right principal eigenvectors  of symmetric $\mathcal{H}$ are the same and have the elements equal to $\frac{1}{\sqrt{N}}$ in the normalized space. Therefore, the variances $\sigma_1$ and $\sigma_2$ of the diagonal elements of $\mathcal{D}^{-1}$ and $\mathcal{D}$, the column sums of $\mathcal{H}$ and their inverses, can somehow describe how much the elements of the principal eigenvector deviate from $\frac{1}{\sqrt{N}}$ which is also the expectation value for randomly generated normalized uniform $N$ numbers. Using these intuitions, we define the estimate $P_{reg_2}$ as:
\begin{equation}
\widetilde{P}_{reg_2}=\frac{\Lambda_1+\Lambda_2}{2}
\end{equation} 
with 
\begin{equation}
\Lambda_1=\frac{\frac{1}{N}-\sigma_1}{\frac{1}{N}+\sigma_1},\ 
\Lambda_2=\frac{\frac{1}{N}-\sigma_2}{\frac{1}{N}+\sigma_2}.
\end{equation}

In addition,  $\alpha_1^2=1$ for stochastic matrices and so for $\mathcal{H}$ when $\mathcal{H}=\mathcal{HD}$. Since the any deviation of the elements of the principal eigenvector  from $\frac{1}{\sqrt{N}}$  would change the equality $\alpha_1^2=1$, we multiply variance by $N$ to get total deviation from 1. Therefore, we define the estimate $\alpha_1^2$ as:
\begin{equation}
\widetilde{\alpha}_1^2=\frac{(1-N\sigma_1)+(1-N\sigma_2)}{2}.
\end{equation}

Note that $\widetilde{\alpha}_1^2$ and $\widetilde{P}_{reg_2}$ are only predictions and may fail to provide good estimates for ${\alpha}_1^2$ and ${P}_{reg_2}$, respectively,  for particularly some structured matrices and matrices whose eigenvectors are close to the columns of an identity matrix. For instance, consider  the following matrix:
\begin{equation}
\left(\begin{matrix}
 21.8214 &  0&    0.6118&    0.4983\\
         0  &14.2944 & 0.4983 &0.6118\\
    0.6118  &  0.4983 &  12.1626&  0\\
    0.4983  & 0.6118  & 0 &5.4111
\end{matrix}\right).
\end{equation}
The eigenvalues of this matrix are
$5.3537, 12.0193,14.4411,$ and  $21.8753$, respectively, associated with the following eigenvectors:
\begin{equation}
\left(
\begin{matrix}
   0.0304613&   0.0597207&   0.0215934&   0.9975166\\
   0.0686662&   0.2074209 & -0.9758165 &  0.0066086\\
  -0.0077623&  -0.9761393  &-0.2076080  & 0.0631720\\
  -0.9971443 &  0.0237068  &-0.0649217   &0.0304360
\end{matrix}\right).
\end{equation}
As computed from the above, the magnitudes of the sums of the components of the eigenvectors  are  very close to each other:  $|-0.90578|,  |-0.68529|,  |-1.22675|,$ and $|1.09773|$, respectively. In addition, the second eigenvector, not associated to the principal eigenvalue, has the largest sum. 
 
 Also note that one can also predict ${\alpha}_1^2$ from the distribution of the matrix elements. For instance \cite{Minc1970maximal,Cioaba2007principal}, 
\begin{equation}
\label{eq:ratioofelements}
max_{ij}\frac{\mu_{1i}}{\mu_{1j}} \leq \frac{max_{ij}h_{ij}}{min_{ij}h_{ij}}, 
\end{equation}
where $h_{ij}$s are the nonzero matrix elements of $\mathcal{H}$.  However, these bounds are generally not tight enough to give good estimates in many cases.

In Sec.\ref{Sec:NumericSimulations}, we shall show the comparisons of these estimates with the computed probabilities for random matrices.
\section{Possible Applications}

In the previous sections, we have showed  the positive eigenpair of an irreducible symmetric non-negative matrix can be obtained by using an initial equal superposition state when the estimated probabilities high. This eliminates  the necessity of knowing an initial approximate eigenvector in the applications of the phase estimation algorithm to the problems involving non-negative matrices. Irreducible non-negative matrices are known to have positive eigenpairs due to Perron-Frobenious theorem. There is also more general but the same statements indicated in this theorem for compact operators \cite{Du2006order}.  One may also apply permutation matrices or  splitting techniques to convert the interested matrix to a non-negative one or to increase the expected probability outcome when the probability is low than an acceptable value.

Since a wide  variety of problems in science can be represented by non-negative matrices,  the  phase estimation algorithm can be applied without an initial approximate eigenvector to these problems. In the following subsections, we shall consider two important classes among these problems: viz., the k-local Hamiltonian problem and the stochastic processes.

\subsection{Applications to Local Hamiltonian Problems}

One of the applications of the non-negative matrices can be found in quantum mechanics where such matrices called stoquastic matrices \cite{Bravyi2008,Bravyi2010}: i.e.,  matrices with non-negative off-diagonal elements.
 If a Hamiltonian operator $\mathcal{H}$, a Hermitian operator acting on $(C^2)^{\otimes n}$, can be expressed as a sum of local Hamiltonians representing the interactions between at most $k$ qubits, then it is called $k$-local $n$-qubit Hamiltonian. More formally, $\mathcal{H}=\sum_sH_s$.  Finding the lowest eigenvalue of $\mathcal{H}$ defines an eigenvalue problem which is so called local Hamiltonian problem. This eigenvalue problem can be also described as a decision problem to decide whether the ground state energy of $H$
is at most $a$  or at least
$b$  for given constants $a$ and $b$ such that $a\leq b$ and $b-a\leq 1/poly(n)$. This problem has been shown to be $QMA$-complete for $k\geq 2$ \cite{Kempe2006complexity} (for k=2, it is QMA-complete only when both negative and positive signs in the Hamiltonian are allowed \cite{Jordan2010}).
 
For the local Hamiltonians with positive off-diagonal matrix elements in the standard basis,  the matrix elements of the corresponding Gibbs density matrix, $\rho = e^{-\beta\mathcal{H}}/Tr(-\beta\mathcal{H})$, are all positive for any $\beta \geq 0$. Due to the theorem given above, 
The ground state energy of $\mathcal{H}$ with positive off-diagonal elements can be associated with the eigenvector whose coefficients are all positive. 
Because one can associate a probability distribution to the ground state, the nature of these Hamiltonians are considered  similar to stochastic processes. Thus, the Hamiltonians of these types are called \textit{stoquastic} \cite{Bravyi2008}.(note that the Hamiltonian is not a stochastic matrix whose rows and columns are sum to one and the principal eigenvalue is one associated with an eigenvector of all ones.)

\subsection{Applications to Stochastic Processes}
The phase estimation algorithm can find the principal eigenpair of a stochastic matrix with probability one \citep{Daskin2014} since the sum of the coefficients is one for the principal eigenvector and zero for the rest of the eigenvectors. However, the stochastic processes are generally defined by non-Hermitian matrices, which are also widely seen in quantum physics: The Hamiltonian of a closed system in equilibrium must be Hermitian to have real energy values. However, nonequilibrium processes and open systems connected to reservoirs can be only described by non-Hermitian models \cite{Efetov1997directed}.

\subsubsection{Simulation of  Non-Hermitian Operators}
 Any non-Hermitian matrix $\mathcal{H}$ can be decomposed into a Hermitian and a skew-Hermitian parts as:
 \begin{equation}
 \label{EqHSH}
 \mathcal{H}=H+S=\frac{1}{2}(\mathcal{H}+\mathcal{H}^\dagger)+\frac{1}{2}(\mathcal{H}-\mathcal{H}^\dagger)
 \end{equation}
 where  $\mathcal{H}^\dagger$ describes the conjugate transpose of $\mathcal{H}$. Here,   
  $H=\frac{1}{2}(\mathcal{H}+\mathcal{H}^\dagger)$ and 
 $S=\frac{1}{2}(\mathcal{H}-\mathcal{H}^\dagger)$ define the nearest Hermitian and skew Hermitian matrices to $\mathcal{H}$, respectively \cite{ClosestHermitian1975}.Therefore, the matrix $H$ can be used as an approximation to $\mathcal{H}$ in the simulation. 
 
  A matrix $\mathcal{H}$ is called normal if $\mathcal{H}^\dagger \mathcal{H}-\mathcal{H}\mathcal{H}^\dagger=0$.
When $\mathcal{H}$ is a normal, it is easy to see that  $H$ and $S$ also commute: $[H,S]=HS-HS=0$. Moreover, because of the symmetry,  all the eigenvalues of $H$ are  real; and because of the skew-symmetry, all the eigenvalues of $S$ involve only  imaginary parts. 
 Since $\mathcal{H}\mathcal{H}^\dagger=\mathcal{H}^\dagger \mathcal{H}$,  the eigenvectors of $H$ and $S$ are the same. In addition, the imaginary part of the eigenvalues of $\mathcal{H}$ are equal to the eigenvalues of $S$ and  the real parts are to the eigenvalues of $H$. 
Hence, using $U_1=e^{iH}$ and $U_2=e^{S}$  in the simulation, one can simulate the non-Hermitian operator $\mathcal{H}$  on quantum computers by using separate two registers to obtain the imaginary  and the real parts of the eigenvalue.
However,   in the stochastic cases, if  $\mathcal{H}$ is normal, it must be also doubly stochastic: i.e., its left and right principal eigenvectors are already known to be a vector of all ones with the eigenvalue one. Therefore, instead of an approximate normal matrix, we can use the closest Hermitian matrix defined above as $H=\frac{1}{2}(\mathcal{H}+\mathcal{H}^\dagger)$. 

In the non-stochastic cases, one can also try to instead find the closest normal matrix  by following the Jacobi algorithm which attempts to diagonalize the matrix by using rotation matrices (Givens rotations). 
and converges to a matrix so called $\Delta H$. 
The best closest matrix in the Frobenius norm  is then defined by the sum of the diagonal elements of $\Delta H$ and the rotation matrices used in the algorithm \cite{Higham89MatrixNearness,Ruhe1987closest}. One can also procure a quantum circuit to implement this closest matrix  by mapping rotation matrices to quantum gates as done in \cite{Vartiainen2004efficient}.
However, since this algorithm is based on the eigenvalue decomposition, the eigenvalues (the diagonal  elements of $\Delta H$, and the eigenvectors (the combination of the rotation matrices ) are already generated for the found normal matrix.

\section{Numerical Results}
\label{Sec:NumericSimulations}
In this section, we compare the estimated $\alpha_1^2$, $P_{reg_1}$, $P_{reg_2}$: i.e. $\widetilde{\alpha}_1^2$, $\widetilde{P}_{reg_1}$, $\widetilde{P}_{reg_2}$, respectively, with the computed probabilities $\alpha_1^2$, $P_{reg_1}$, $P_{reg_2}$ for the randomly generated matrices described below. 
\subsection{Random Symmetric Matrices with Non-Negative Off-Diagonal Elements}
In MATLAB, we generate a random  symmetric matrix with non-negative off-diagonal elements as follows: first a random diagonal matrix and a strictly upper triangular sparse non-negative matrix are generated. Then, these matrices are combined to have a symmetric matrix with non-negative off-diagonal elements:
\begin{center}
\begin{verbatim}
%MATLAB code to generate a random symmetric matrix
% with nonnegative off-diagonal elements
R = triu(sprand(N,N,0.5),1); % N is the dimension, 
% and 0.5 is the sparsity of the matrix. 
L = randn(N,1);
H = R + R' + diag(L);
\end{verbatim}
\end{center}
Fig.\ref{fig:randomMatrices} shows the comparisons of the estimated and computed  probabilities for matrices of different dimensions generated by following the above code. As shown in the figure, the estimated and computed probabilities are almost the same and very high (close to 1). Furthermore, the success probability, $P_{reg_1}$, with the Hadamard gates applied to \ket{reg_2} in the end is almost one and higher than the probability, $\alpha_1^2$, without the Hadamard gates.
\subsection{Random 3-Local Hamiltonians}

\begin{figure*}[h]
\centering
\subfigure[512 by 512 3-local Hamiltonians]{
\includegraphics[width=0.45\textwidth]{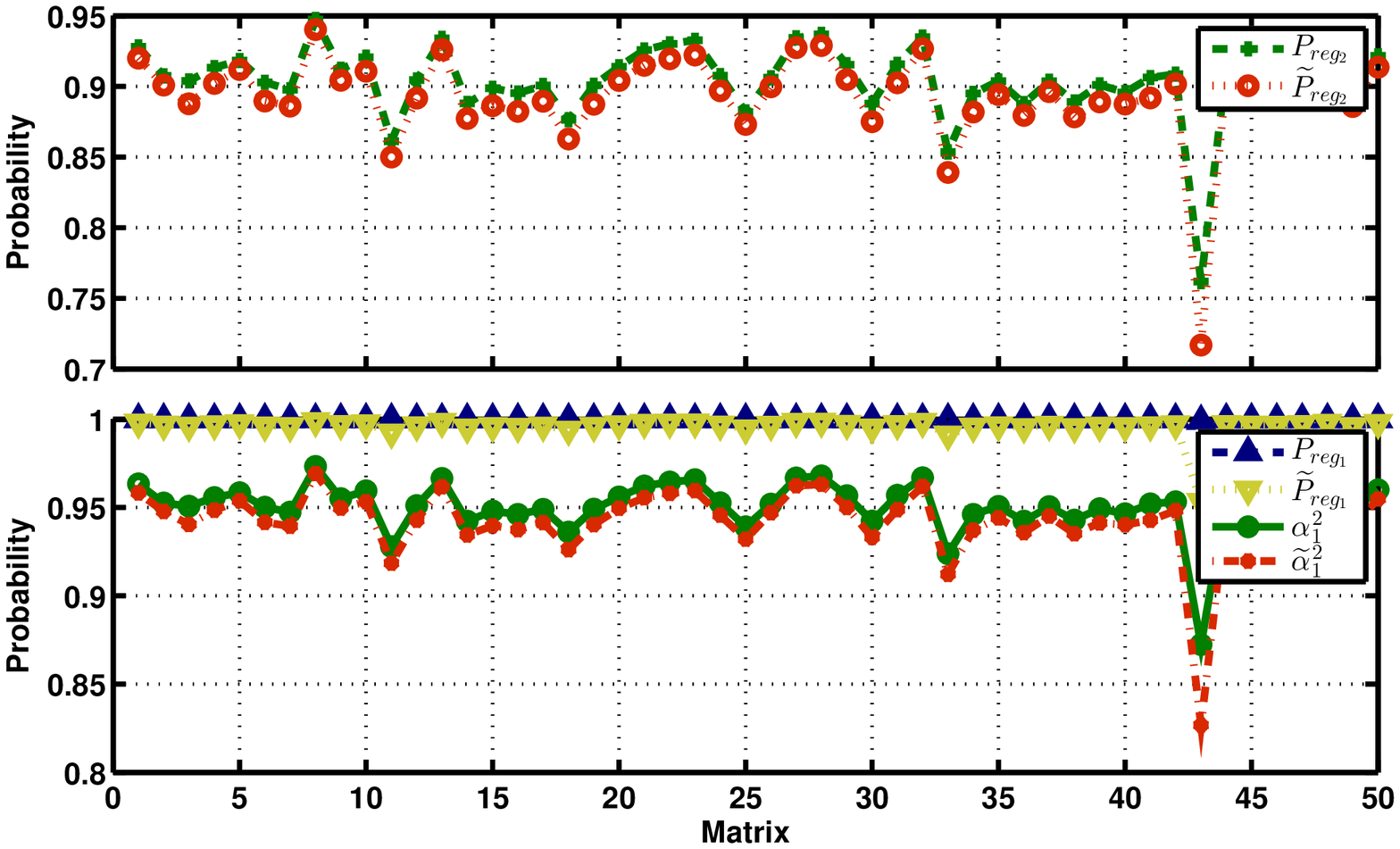}}
\subfigure[1024 by 1024 3-local Hamiltonians]{
\includegraphics[width=0.45\textwidth]{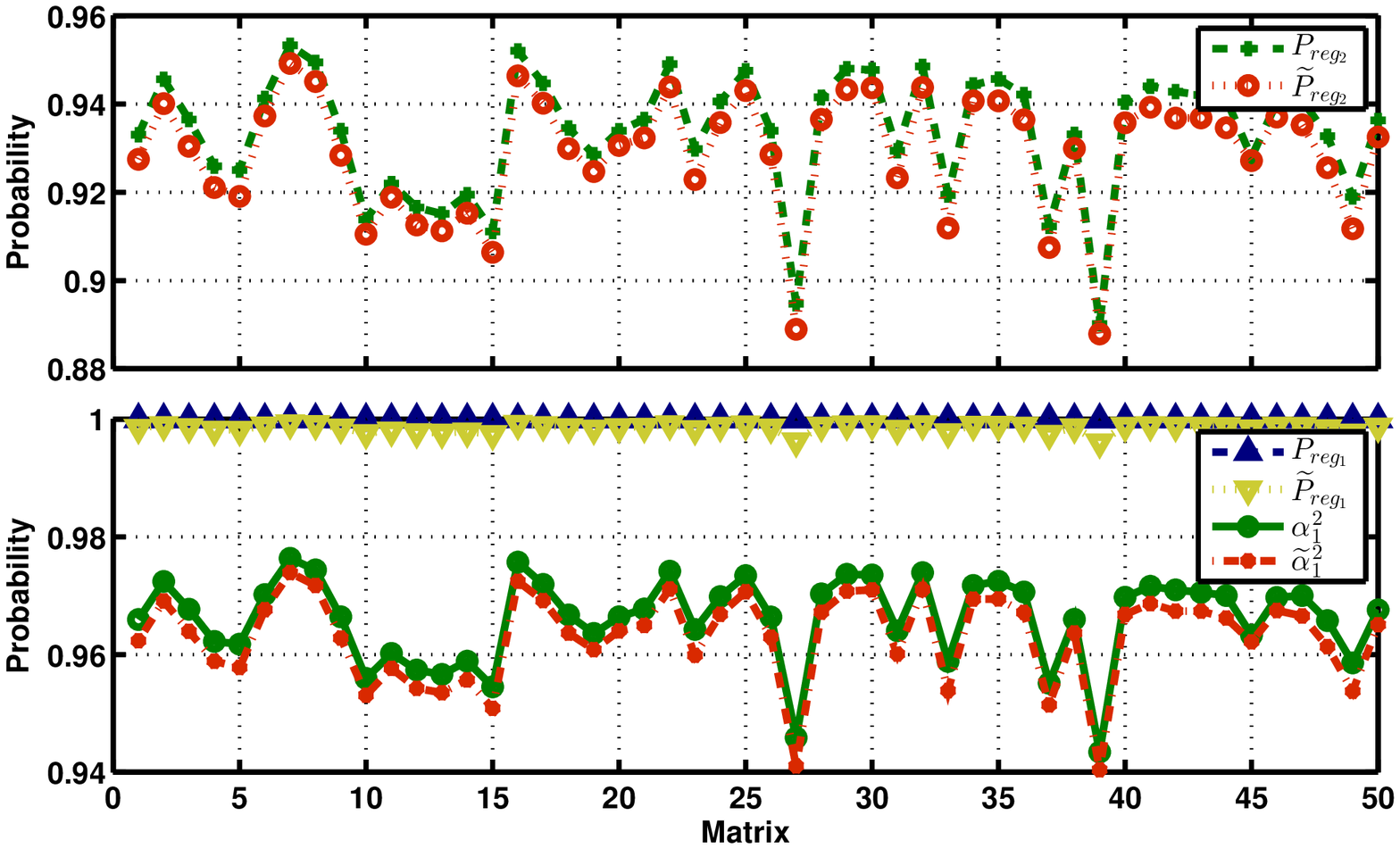}}
\subfigure[2048 by 2048 3-local Hamiltonians]{
\includegraphics[width=0.45\textwidth]{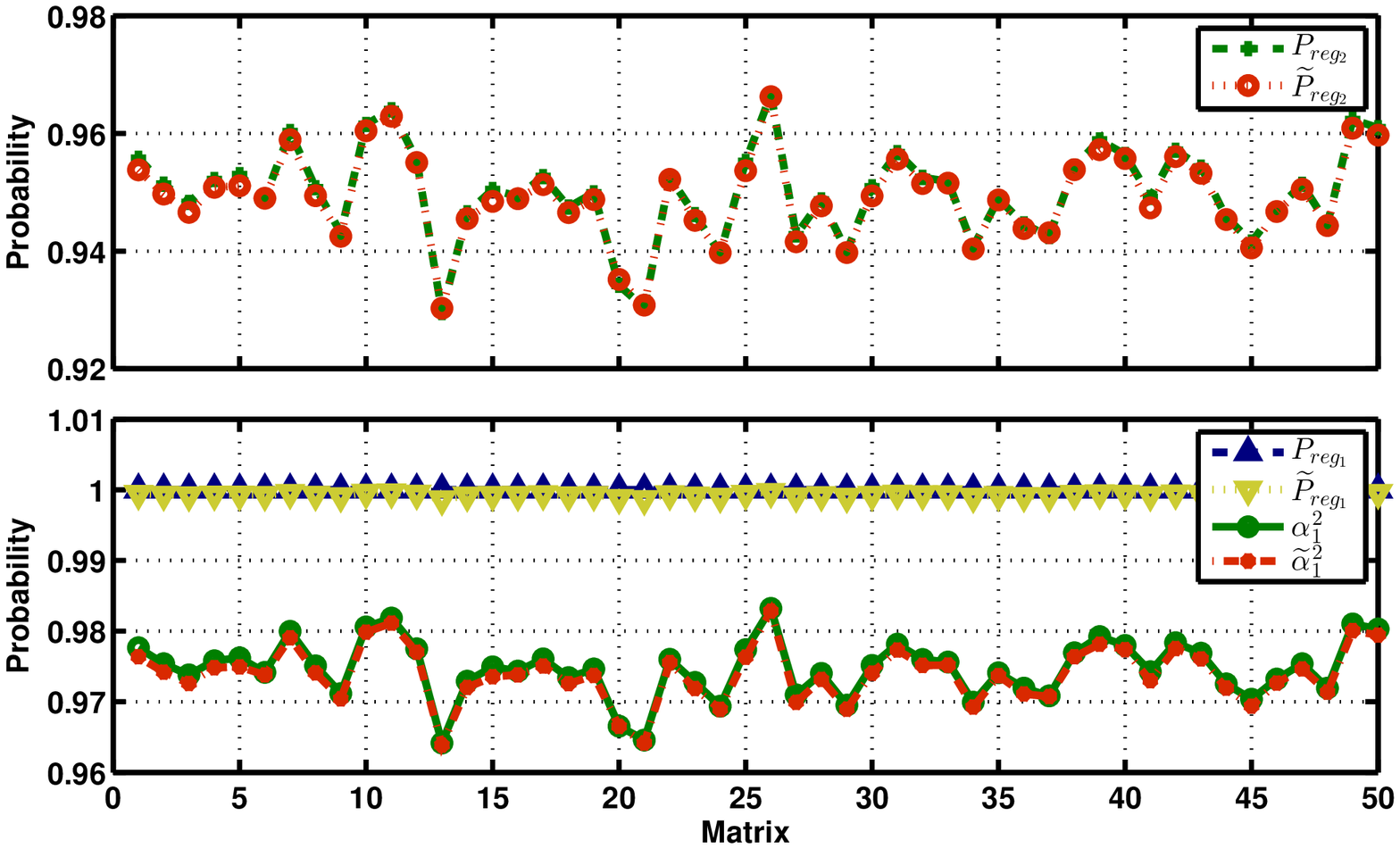}}
\subfigure[4096 by 4096 3-local Hamiltonians]{
\includegraphics[width=0.45\textwidth]{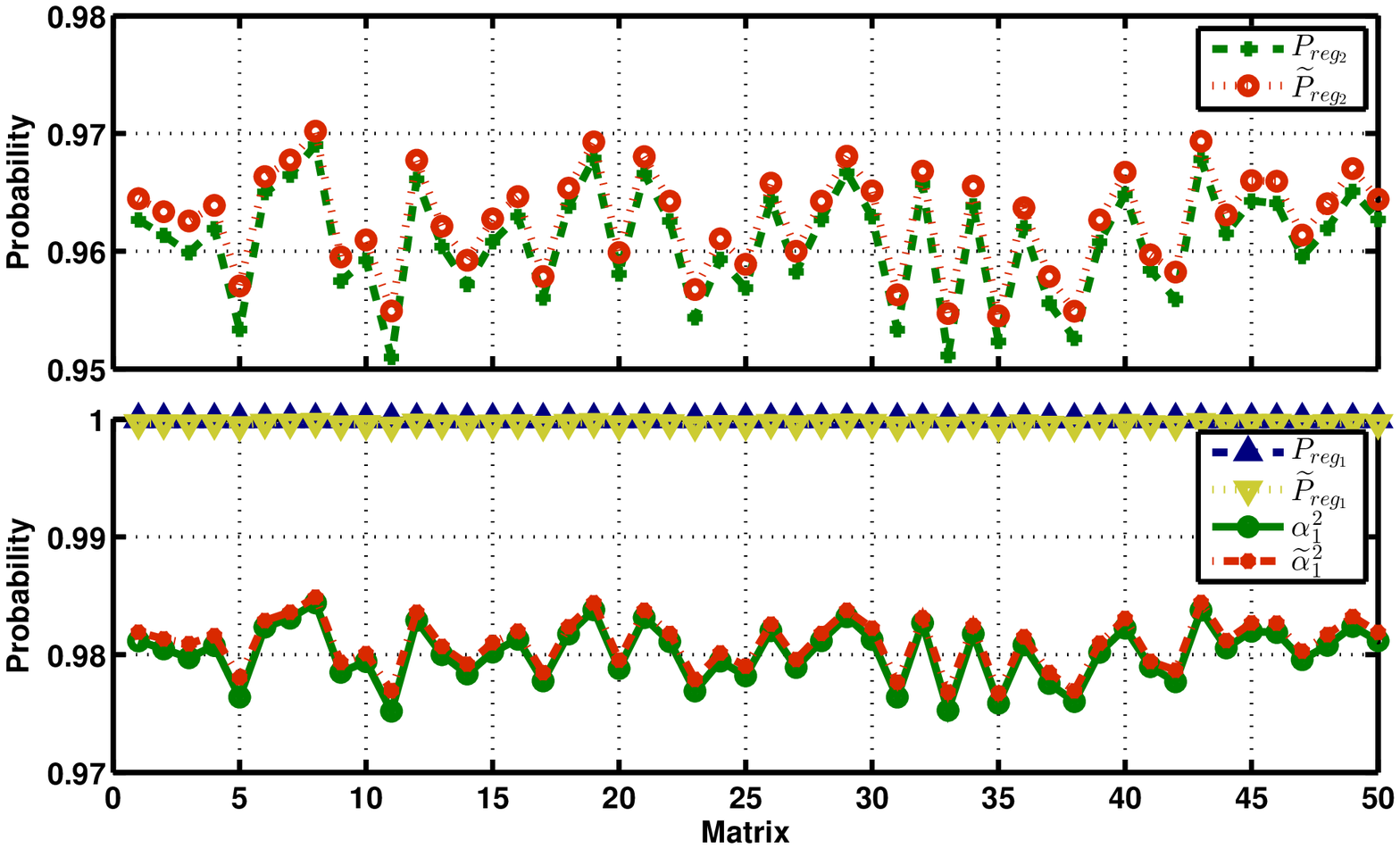}}
\caption{\label{fig:randomLocalH1} Representations  of the comparisons of the estimated probabilities $\widetilde{\alpha}_1^2$, $\widetilde{P}_{reg_1}=\frac{\widetilde{\alpha}_1^4}{\widetilde{P}_{reg_2}}$, $\widetilde{P}_{reg_2}$ with the computed probabilities $\alpha_1^2$, $P_{reg_1}=\frac{\alpha_1^4}{P_{reg_2}}$, and $P_{reg_2}$  for random 3-local Hamiltonians of different dimensions described in Eq.(\ref{Eq:LocalH1}).}
\end{figure*}

As our first example, the following Hamiltonian is employed:
\begin{equation}
\label{Eq:LocalH1}
\mathcal{H}_{XZ}=\sum_{i,j,k}K_{ijk}X_iX_jX_k+J_{ijk}Z_iZ_jZ_k
\end{equation}
with $0 \leq K_{ijk}\leq 1$ and $-1 \leq J_{ijk} \leq 1$. Here, $X$ and $Z$ are the Pauli spin operators. Choosing $K_{ijk}$ and $J_{ijk}$ randomly, 50 numbers of random matrices are generated for each 9, 10, 11, 12, and 13 qubit systems and show the estimated and computed probabilities in Fig.\ref{fig:randomLocalH1}.

\begin{figure*}[h]
\centering
\subfigure[512 by 512 Matrices]{
\includegraphics[width=0.45\textwidth]{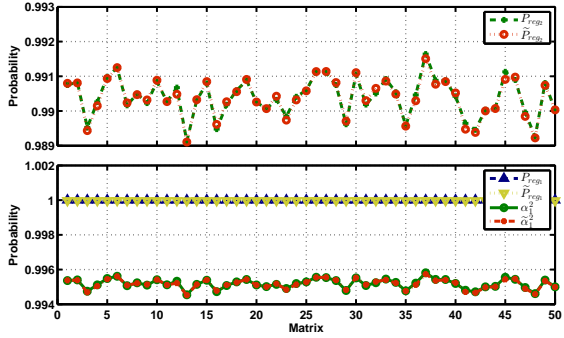}}
\subfigure[1024 by 1024 Matrices]{
\includegraphics[width=0.45\textwidth]{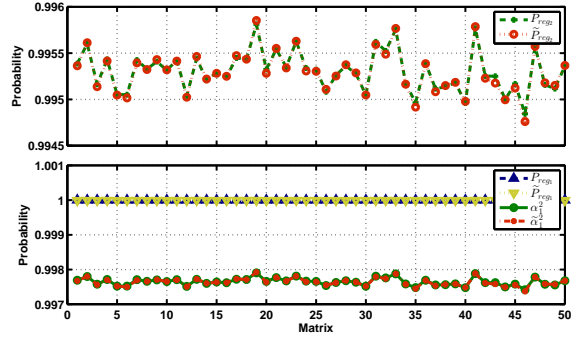}}
\subfigure[2048 by 2048 Matrices]{
\includegraphics[width=0.45\textwidth]{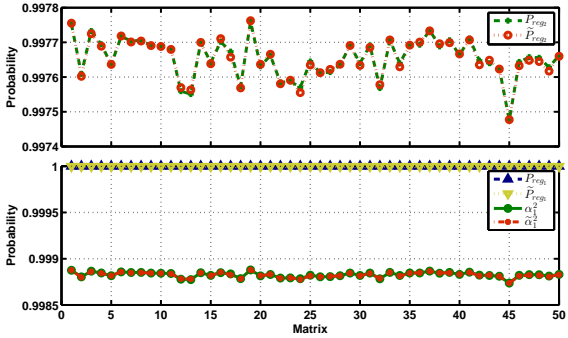}}
\subfigure[4096 by 4096 Matrices]{
\includegraphics[width=0.45\textwidth]{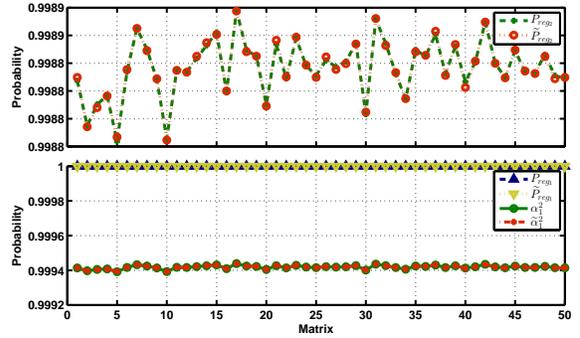}}
\caption{\label{fig:randomMatrices} Representations  of the comparisons of the estimated probabilities $\widetilde{\alpha}_1^2$, $\widetilde{P}_{reg_1}=\frac{\widetilde{\alpha}_1^4}{\widetilde{P}_{reg_2}}$, $\widetilde{P}_{reg_2}$ with the computed probabilities $\alpha_1^2$, $P_{reg_1}=\frac{\alpha_1^4}{P_{reg_2}}$, and $P_{reg_2}$  for random non-negative symmetric matrices of different dimensions.  }
\end{figure*}

\begin{figure*}[h]
\centering
\subfigure[512 by 512 3-local Hamiltonians]{
\includegraphics[width=0.45\textwidth]{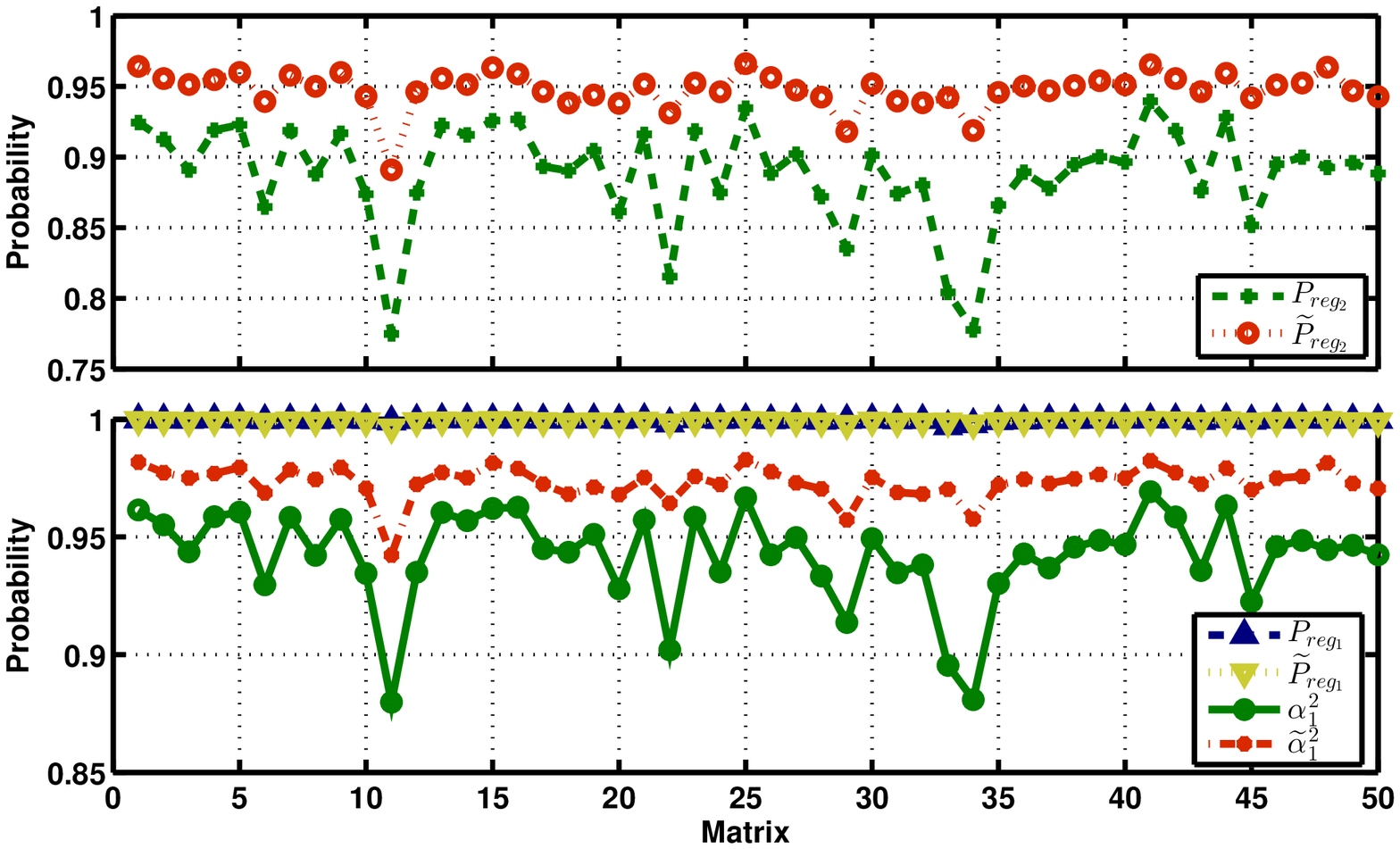}}
\subfigure[1024 by 1024 3-local Hamiltonians]{
\includegraphics[width=0.45\textwidth]{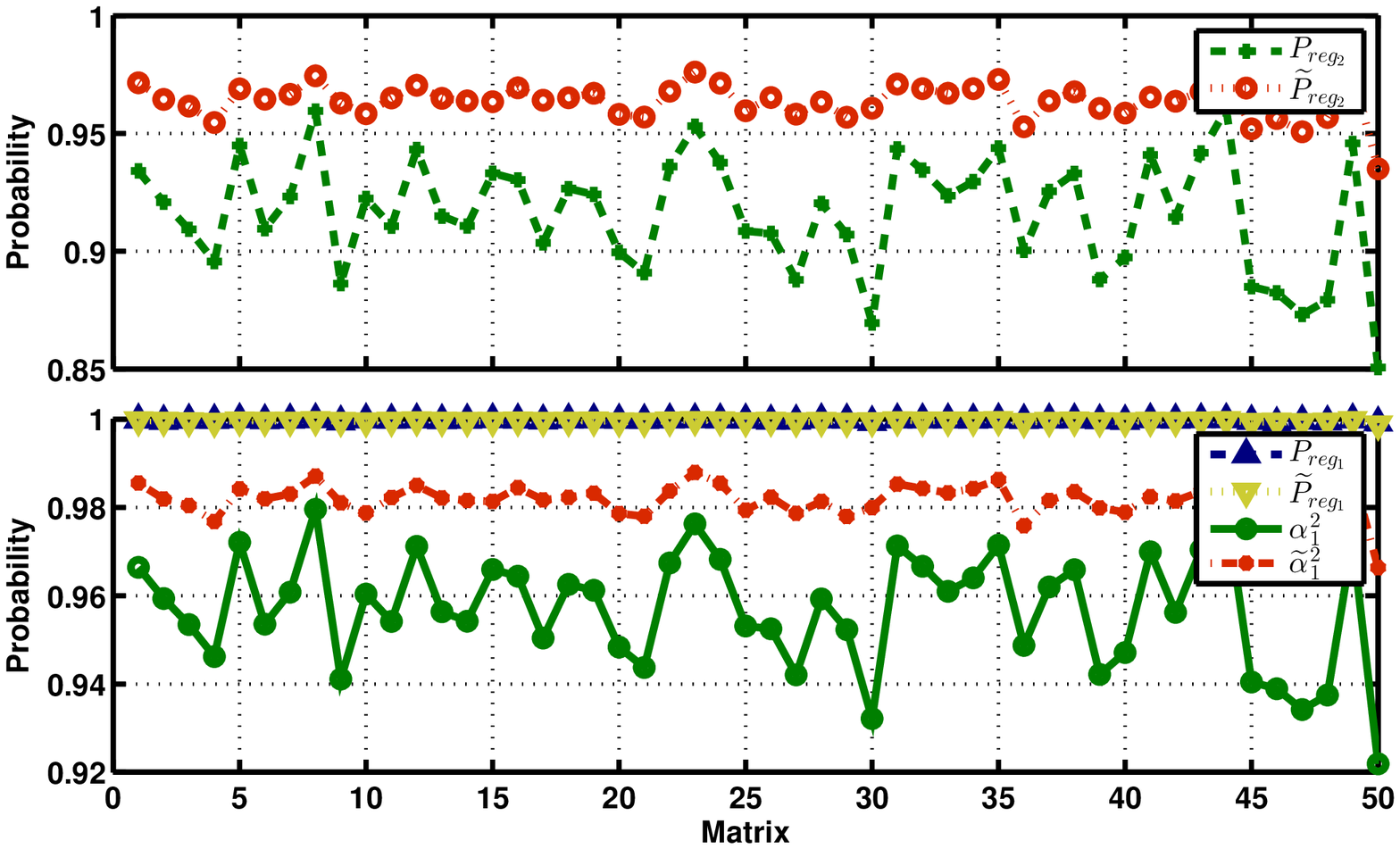}}
\subfigure[2048 by 2048 3-local Hamiltonians]{
\includegraphics[width=0.45\textwidth]{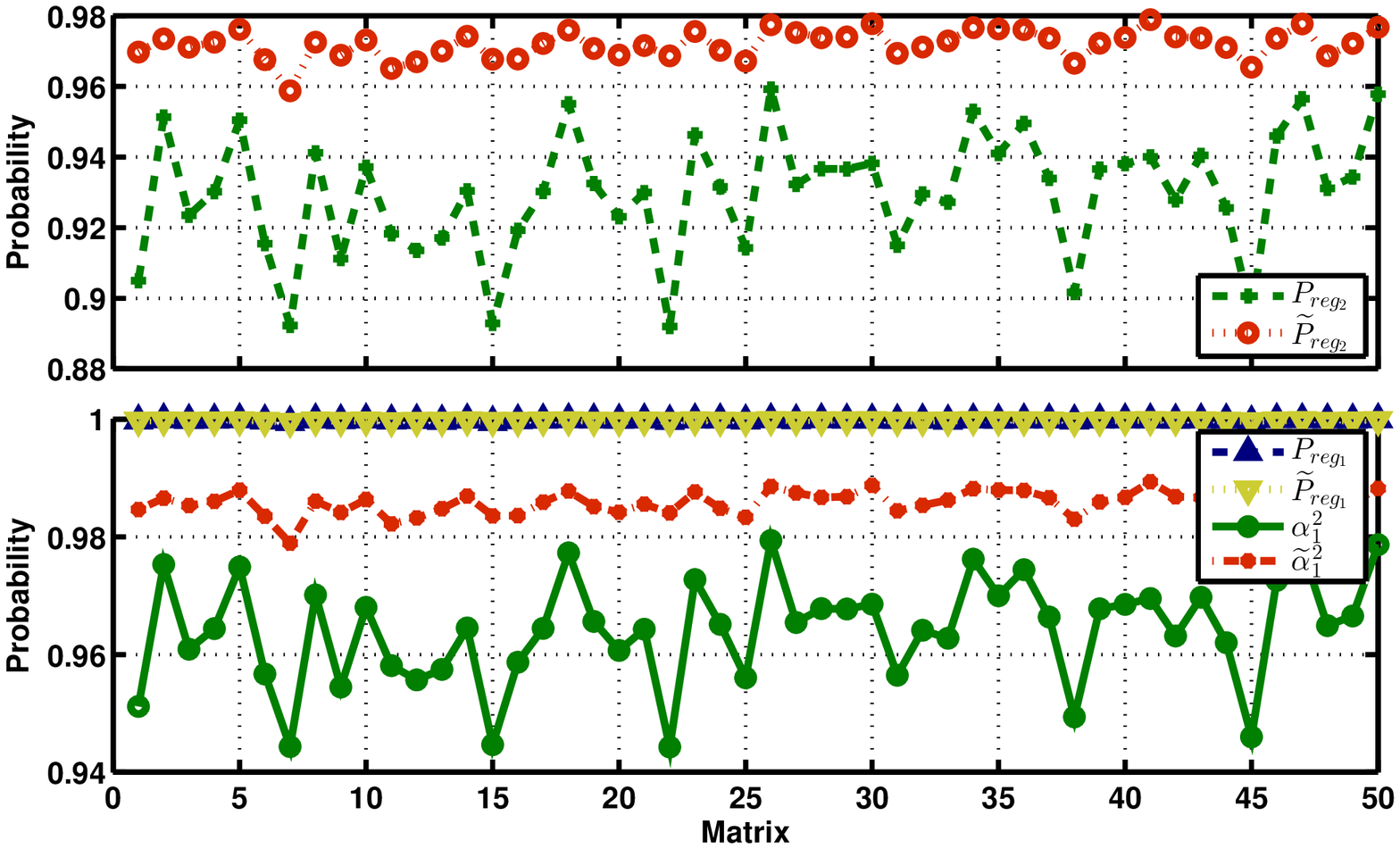}}
\subfigure[4096 by 4096 3-local Hamiltonians]{
\includegraphics[width=0.45\textwidth]{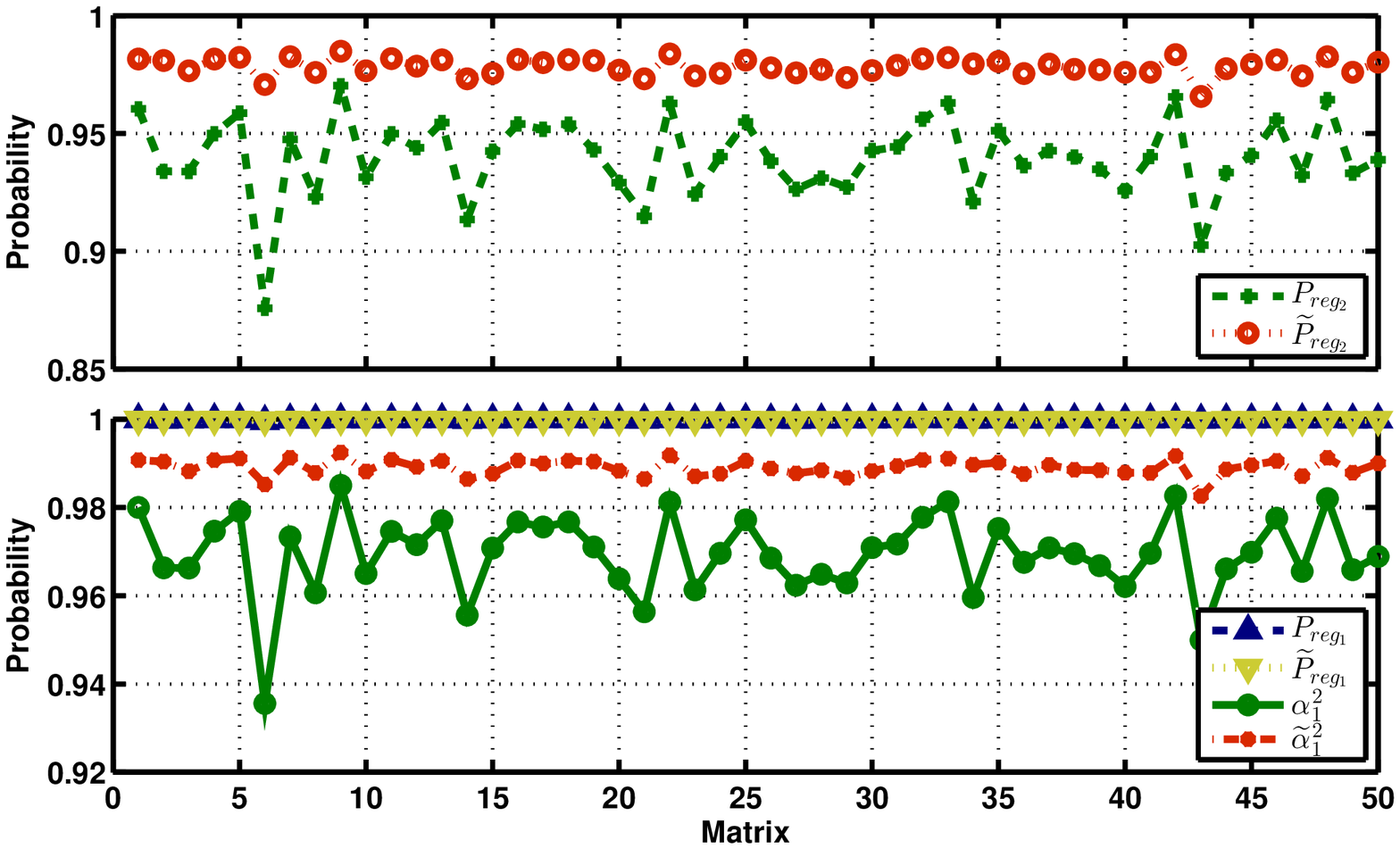}}
\caption{\label{fig:randomLocalH2} Representations  of the comparisons of the estimated probabilities $\widetilde{\alpha}_1^2$, $\widetilde{P}_{reg_1}=\frac{\widetilde{\alpha}_1^4}{\widetilde{P}_{reg_2}}$, $\widetilde{P}_{reg_2}$ with the computed probabilities $\alpha_1^2$, $P_{reg_1}=\frac{\alpha_1^4}{P_{reg_2}}$, and $P_{reg_2}$  for random 3-local Hamiltonians of different dimensions described in Eq.(\ref{Eq:LocalH2}).}
\end{figure*}

In the second example, 1-body and 2-body interaction terms are also  included separately in the Hamiltonian:
\begin{equation}
\label{Eq:LocalH2}
\begin{split}
\mathcal{H}_{XZ}& =\sum_{i}d_iX_i+h_iZ_i+\sum_{i,j}{K_1}_{ij}X_iX_j+
{J_1}_{ij}Z_iZ_j\\ &+\sum_{i,j,k}{K_2}_{ijk}X_iX_jX_k+
{J_2}_{ijk}Z_iZ_jZ_k
\end{split}
\end{equation}
with $0 \leq d_i,{K_1}_{ijk},{K_2}_{ijk}\leq 1$ and $-1 \leq h_i,{J_1}_{ijk},{J_2}_{ijk} \leq 1$.
As done for Eq.(\ref{Eq:LocalH1}, choosing the coefficients randomly, Hamiltonians of different sizes are generated. The results are shown in Fig.\ref{fig:randomLocalH2}. As seen in the figure, the estimation is not as good as the cases in Fig.\ref{fig:randomMatrices} and Fig.\ref{fig:randomLocalH1}. This is because  the ratio defined in Eq.(\ref{eq:ratioofelements}) is generally higher for the matrices generated from Eq.(\ref{Eq:LocalH2}) in comparison to the ones from Eq.(\ref{Eq:LocalH1}).
\section{Conclusion}
In this paper, we have showed  the positive eigenpair of an irreducible symmetric non-negative matrix can be obtained when the estimated probabilities high.  This eliminates  the necessity of knowing an initial approximate eigenvector in the applications of the phase estimation algorithm to a wide variety of problems involving non-negative matrices. Here, we also have discussed how to apply it to local Hamiltonian problems and stochastic processes and showed the comparisons of the estimated success probabilities with the computed ones for random symmetric matrices and 3-local Hamiltonians.

\end{document}